# High-Pressure Synthesis and Superconductivity of Ag-doped Topological Crystalline Insulator SnTe ($Sn_{1-x}Ag_xTe$ with $x$ = 0-0.5)

Yoshikazu Mizuguchi[1], and Osuke Miura[1]

[1]*Tokyo Metropolitan University, 1-1, Minami-osawa, Hachioji 192-0397*

We have synthesized the single-phase polycrystalline samples of $Sn_{1-x}Ag_xTe$, Ag-doped topological crystalline insulator SnTe, with a range of $x \leq 0.5$ using a high-pressure synthesis method. The crystal structure of $Sn_{1-x}Ag_xTe$ at room temperature is a cubic NaCl-type structure, which does not vary upon Ag substitution. Bulk superconductivity with a transition temperature of 2.4 K was observed for $x$ = 0.15-0.25, and the optimal Ag content was $x$ = 0.2. The $Sn_{1-x}Ag_xTe$ superconducting phase will be useful for understanding the superconductivity nature and mechanisms of the carrier-doped SnTe system.

Topological insulators are a new class of materials that exhibits fully-gaped states in bulk and gapless metallic states at the surface, which are protected by time-reversal symmetry [1-3]. In addition, superconductors derived from topological insulators have been actively studied with expectations of the emergence of Majorana fermions [4-8]. Recently, SnTe has been drawing much attention as a topological crystalline insulator, in which the topological phase is protected by crystal symmetries [5,9-11]. SnTe itself shows superconductivity with a transition temperature ($T_c$) up to 0.3 K with hole concentration of $10^{21}$ $cm^{-3}$ generated by Sn defects ($Sn_{1-x}Te$) [12,13], and the $T_c$ increases up to 4.8 K by partial substitutions of Sn by In [13-16]. Recently, Haldolaarachchige *et al.* reported that the In in $Sn_{1-x}In_xTe$ does not act as a trivial hole dopant at high In contents [17]. The valence of In cannot be regarded as simple $In^+$, and the sign of Hall coefficient changes to negative in the In-rich region. Therefore, to understand the nature of the superconductivity in carrier-doped SnTe and discuss the relationship between the superconductivity and the topological crystalline insulator states, further investigations with the other SnTe-based superconductors are necessary.

Similarly to the In substitution, Ag substitution for the Sn site can dope holes (assuming $Ag^+$ valence), while the solubility limit of Ag for the Sn site is very low [12]. However, trace of superconductivity in $Sn_{0.97-x}Ag_xTe$ was previously observed in Ref. 12. According to the previous report, homogeneous (annealed) $Sn_{0.97-x}Ag_xTe$ ($x$ = 0.1-0.14) showed superconductivity below 1 K, while the $T_c$ of the as-cast sample was ~2K; the $T_c$ was measured by magnetic susceptibility (arb. unit) measurement. This facts indicate that the observation of superconductivity ($T_c$ ~ 2 K) would be resulting from inhomogeneity of the Ag content, and one can expect that bulk superconductivity will be achieved at higher $x$ in $Sn_{1-x}Ag_xTe$. Unfortunately, the $Sn_{1-x}Ag_xTe$ samples prepared at ambient pressure exhibit *phase separation* due to the low solubility limit [12]. We also confirmed that the $Sn_{1-x}Ag_xTe$ samples for $x$ (nominal) = 0.3, 0.4, and 0.5 are mixed-phase samples composed of the phase1 (Sn-rich phase) and phase2 (nearly $Sn_{0.5}Ag_{0.5}Te$ phase) when the samples were prepared by using conventional solid-state reaction at ambient pressure condition (reaction in an evacuated quartz tube). To suppress the phase separation into two stable phases and to obtain the intermediate region of $0.2 \leq x \leq 0.5$, we utilized the high-pressure (HP) synthesis technique, and successfully obtained single-phase samples of $Sn_{1-x}Ag_xTe$ with $0 \leq x \leq 0.5$. As expected, the HP-$Sn_{1-x}Ag_xTe$ samples with $x$ = 0.15-0.25 are bulk superconductors with $T_c$ of 2.4 K. Here, we show the synthesis (using both conventional solid state reaction and high-pressure methods), crystal structure, and superconducting properties of $Sn_{1-x}Ag_xTe$.

In this study, we prepared two kinds of $Sn_{1-x}Ag_xTe$ polycrystalline samples with $x$ (nominal) = 0-0.5, which are denoted as AP- and HP-$Sn_{1-x}Ag_xTe$. The AP samples were prepared by a conventional solid-state-reaction method. The Ag (99.9%) and Sn (99.99%) powders and Te (99.9999%) grains with a nominal composition of $Sn_{1-x}Ag_xTe$ were mixed, sealed in an evacuated quartz tube, and heated at 600ºC for 10 hours. The non-doped SnTe sample was prepared by melting the mixture of Sn and Te at 900ºC. The HP samples were prepared using a high-pressure (HP) synthesis method with a cubic-anvil-type instrument (CT Factory). The Ag, Sn, and Te powders with a nominal composition of $Sn_{1-x}Ag_xTe$ were mixed, pressed into a pellet with a diameter of 5 mm. The pellet was placed in a BN capsule, and the sample was heated using a carbon-capsule heater at ~500ºC for 30 minutes under a pressure of 2 GPa.

The obtained samples were characterized by the $\theta$-$2\theta$ method using an X-ray diffractometer (Miniflex-600 Dtex-Ultra, RIGAKU) equipped with a Cu-K$\alpha$ radiation at room temperature. The lattice constant ($a$) was calculated by the Rietveld method using RIETAN-FP software [18]. Magnetization measurements were performed using a superconducting interference device (SQUID) magnetometer with a VSM mode (MPMS-3, Quantum design). Temperature dependence of magnetic susceptibility ($\chi$) was measured after zero-field cooling (ZFC) and field cooling (FC). Temperature dependence of electrical resistivity ($\rho$) was measured using a four-probe method with a GM refrigerator system (Thermal Block). The current used in the resistivity measurement was 2 mA. The Au electrodes were put on the sample using silver pastes. The actual composition of the samples were analyzed using energy dispersive X-ray spectroscopy (EDX) (TM3030-SwiftED, Hitachi).

Figure 1(a) shows the X-ray diffraction (XRD) patterns for AP-$Sn_{1-x}Ag_xTe$. For $x$ = 0-0.2, all the peaks were explained using the NaCl-type structure model ($Fm$-3$m$, #225, $O_h^5$). For $x$ = 0.3-0.5, phase separation into two phases (phase1 and phase2) with different lattice constants were observed. The calculated lattice constants are summarized in Table I and plotted in Fig. 2 as a function of $x$. The lattice constant basically tends to decrease with increasing Ag content. However, the shrinkage saturates at above $x = 0.2$, suggesting that the Ag is not expectedly substituted for the Sn site at this region. As a fact, the phase2 arises and its peak intensity is enhanced with increasing $x$; hence, single-phase samples with a higher Ag content cannot be obtained. These results are consistent with the previous studies on $Sn_{0.97-x}Ag_xTe$ [12].

Figure 1(b) shows the XRD patterns for HP-$Sn_{1-x}Ag_xTe$. For $x = 0.2$-$0.5$, the all the peaks were explained using the NaCl-type structure model, while the $x = 0.15$ sample contained a minor phase with a Sn-rich composition. Namely, the single-phase samples of $Sn_{1-x}Ag_xTe$ can be prepared by utilizing HP synthesis technique. As shown in Fig. 2, the lattice constant monotonically decreases up to $x = 0.33$. Notably, the lattice constants for $x = 0.2$ for the AP and the HP samples clearly differ from to each other. This suggests that the AP-$Sn_{0.8}Ag_{0.2}Te$ can be regarded as a single phase within the resolution of the laboratory XRD, but the actual Ag content should be less than the nominal value or the structure is disordered; in fact, EDX analysis indicated that the (Sn/Ag):Te ratio is randomly deviated from 50:50 in the AP-$Sn_{0.8}Ag_{0.2}Te$. One can find that the lattice constant of HP-$Sn_{0.5}Ag_{0.5}Te$ ($x = 0.5$) almost corresponds to those of the phase2 in the AP samples, suggesting that the $Sn_{0.5}Ag_{0.5}Te$ phase is stable probably due to decreased disorder of the Sn/Ag site and/or the valence of Sn and/or Ag. Figure 3 shows the fitting result of the Rietveld refinement for the XRD pattern of HP-$Sn_{0.8}Ag_{0.2}Te$. The resulted reliability factor was 6.5%. The profile at higher angles are nicely fitted as well, indicating that the NaCl-type structure model is correct for this phase. However, we have to mention that the refinement of the Ag content at the Sn site cannot be achieved due to their close atomic number. The atomic number of Te is also close to that of Ag and Sn; hence the simple cubic structure may be possible [19]. However, our compositional analysis of the HP-$Sn_{0.8}Ag_{0.2}Te$ sample revealed that the ratio of (Sn/Ag) and Te is 50:50 with an error of ±2% at whole the sample surface. This implies that the Te site is ordered from the other metal site (Sn/Ag site) because the ratio can be randomly distributed in the simple cubic structure. The analyzed Ag content for the single-phase samples is plotted in the inset of Fig. 2.

Let us briefly discuss the reason for the emergence of phase separation in the AP samples. As mentioned in the introduction, in the $Sn_{1-x}In_xTe$ superconductor family, the solubility limit of In for the Sn site is about 50% [16], while the solubility limit of Ag for the Sn site in AP-$Sn_{1-x}Ag_xTe$ is very low (less than 20%). The difference in these two SnTe-based systems is the evolution of the lattice constant. The lattice volume of SnTe slightly decreases by the In substitution, and it largely decreases by the Ag substitution. For example, the lattice constants of $Sn_{0.5}In_{0.5}Te$ [17] and HP-$Sn_{0.5}Ag_{0.5}Te$ are 6.265 and 6.083 Å, respectively. To discuss the lattice volume evolution, comparing ionic radius is useful. The ionic radius of $Sn^{2+}$, $Sn^{4+}$, $In^{+}$, and $In^{3+}$ is 93, 74, 132, and 92 pm, respectively. The evolution of lattice constant with increasing In content can be understood by these values. The ionic radius of $Ag^+$ and $Ag^{2+}$ is 113 and 89 pm, respectively. The radius of Ag is comparably smaller than that of In. In particular, the radius of $Ag^{2+}$ is clearly smaller than that of $Sn^{2+}$ and $In^{3+}$. Therefore, on the basis of the lattice volume evolution and the possible ionic radius ($Ag^+$, $Ag^{2+}$ or the mixed valence of them), we assume that the valence of Ag may be not simply $Ag^+$ but mixed state with $Ag^{2+}$. In addition, we note that the lattice constant of the phase1 in the AP samples is close to that of SnTe, implying the possibility of the simple $Ag^+$ valence state. In contrast, the lattice constant of the phase2 is clearly smaller than those of the phase1 and SnTe. This fact suggests that the Ag valence is possibly $Ag^{2+}$ in the phase2. This scenario proposes that the phase separation occurs due to the valence separation of $Ag^+$ and $Ag^{2+}$. Therefore, the precise determination of the valence state of Ag and Sn is needed to understand the crystal structure evolution. In general, the HP synthesis technique is powerful skill in synthesizing a compressed phase as demonstrated in several superconductors [19-23]. We consider that the $Sn_{1-x}Ag_xTe$ phases with $x$ up to 0.5 were stabilized under high pressure environment. In addition, the HP environment may stabilize the mixed valence state of $Ag^+$ and $Ag^{2+}$ in $Sn_{1-x}Ag_xTe$. Our results on the HP synthesis of $Sn_{1-x}Ag_xTe$ propose that the HP synthesis technique should be useful for exploring new SnTe-based or other NaCl-type superconductors with mixed valence states.

We firstly show the superconducting properties of the AP samples. Figure 4(a) shows the temperature dependences of $4\pi\chi$ (ZFC) for AP-$Sn_{1-x}Ag_xTe$. For $x$ = 0.3-0.5, weak diamagnetic signals detecting the evolution of superconducting states are observed below 2 K, but the shielding volume fractions (less than 5% at 1.8 K) are very low as a bulk superconductor. On the basis of the present susceptibility measurements and the previous reports [12], the superconductivity observed in these AP-$Sn_{1-x}Ag_xTe$ is not bulk in nature at least above 1.8 K.

In contrast to the AP samples, the HP-$Sn_{1-x}Ag_xTe$ samples show large diamagnetic signals. Figure 4(b) shows the temperature dependences of $4\pi\chi$ (ZFC) for HP-$Sn_{1-x}Ag_xTe$. Bulk superconductivity is observed for $x$ = 0.15-0.25; here, we confirmed bulk superconductivity states with a criterion of $4\pi\chi$(1.8K-ZFC) < -0.5 (50% shielding volume fraction). The highest $T_c$ ($T_c^{onset}$ = 2.4 K) and the largest shielding volume fraction (nearly 100%) are obtained for $x$ = 0.2. In addition, the XRD profile for the HP-$Sn_{0.8}Ag_{0.2}Te$ sample was nicely refined by the Rietveld analysis as shown in Fig. 2. Therefore, we concluded that the optimal Ag content is $x$ = 0.2 for the HP-$Sn_{1-x}Ag_xTe$ system.

We show the superconducting properties of HP-$Sn_{0.8}Ag_{0.2}Te$ in detail. Figure 5(a) shows the temperature dependence of $4\pi\chi$ (ZFC and FC) for HP-$Sn_{0.8}Ag_{0.2}Te$, and the inset shows the magnetization-field characteristics (at 1.8 K) at low magnetic fields. These magnetic properties suggest that the HP-$Sn_{0.8}Ag_{0.2}Te$ superconductor is a type-II superconductor. To evaluate the upper critical field ($\mu_0 H_{c2}$), the temperature dependences of $\chi$ were measured at various magnetic fields up to 0.2 T. Since superconductivity is suppressed with increasing magnetic field, we defined the magnetic $T_c$ as the temperature at which $\chi$ became negative as shown in the inset of Fig. 5(b), which is an example of the estimation of the magnetic $T_c$ of HP-$Sn_{0.8}Ag_{0.2}Te$ at 0.2 T. The estimated magnetic $T_c$ and the magnetic field are plotted in Fig. 5(b). The dashed line is a linear fitting of the $\mu_0 H_{c2}$ ($T$), which gives 1.04 T at 0 K. Using the WHH model [24], we estimated the $\mu_0 H_{c2}$ (0) was 0.72 T for HP-$Sn_{0.8}Ag_{0.2}Te$. Figure 6 shows the temperature dependence of electrical resistivity ($\rho$-$T$) for HP-$Sn_{0.8}Ag_{0.2}Te$ below the room temperature. Metallic behavior is observed at whole temperatures down to $T_c$. The inset shows the enlarged $\rho$-$T$ at around $T_c$. The transport $T_c^{onset}$ and $T_c^{zero}$ are 2.4 and 2.0 K, respectively.

To discuss the evolution of bulk superconductivity in Ag-doped SnTe, we plotted the magnetic $T_c^{onset}$ and $T_c^{5\%(4\pi\chi)}$ as a function of lattice constant ($a$) in Fig. 7. The $T_c^{5\%(4\pi\chi)}$ was defined to be the temperature where the magnetic shielding volume fraction exceeded 5%, which can be a good scale (criterion) for discussing the emergence of bulk superconducting states in the present case. The criterion excludes the superconducting transitions observed in the AP-$Sn_{1-x}Ag_xTe$ samples from bulk superconductivity. In addition, the values of $-4\pi\chi$(ZFC) at 1.8 K are also plotted in Fig. 7. Bulk superconductivity is observed in a region of $a$ = 6.11-6.22 Å. The $T_c^{onset}$ does not depend on the lattice constant (or Ag content). This may be resulting from inhomogeneity of the Ag content or local structure in the sample. The presence of local regions with the optimal lattice constant (or Ag content) can result in the onset of superconductivity. In contrast, the $T_c^{5\%(4\pi\chi)}$ and the $-4\pi\chi$(ZFC) plot shows a dome-shaped phase diagram. In addition, the evolution of $T_c^{5\%(4\pi\chi)}$ corresponds to the magnitude of the shielding volume fraction at 1.8 K. Bulk superconductivity disappears at $a$ = 6.08 ($x$ = 0.5). In the case of In-substituted $Sn_{1-x}In_xTe$, 50%-In substitution resulted in the highest $T_c$. However, superconductivity was not observed for $x$ = 0.5 at least above 1.8 K in $Sn_{0.5}Ag_{0.5}Te$. The difference in the emergence of superconductivity in In- and Ag-substituted SnTe may be due to the difference of the valence state of the dopant element (In or Ag). In addition, as mentioned in the XRD part, we assume that the crystal structure of $Sn_{0.5}Ag_{0.5}Te$ ($x$ = 0.5) can contain some ordering of the Sn/Ag site and/or Sn and/or Ag valence. The Sn/Ag site ordering may cause the structure distortion into a lower symmetry phase as observed in $Ag_{0.5}Bi_{0.5}Se$ ($AgBiSe_2$), whose crystal structure changes from cubic to hexagonal at lower temperatures [25]. Therefore, detailed analysis of the valence states (for both Ag and Sn) and low-temperature crystal structure of $Sn_{1-x}Ag_xTe$ is important to further understand the superconductivity phase diagram of the $Sn_{1-x}Ag_xTe$ system.

In conclusion, we have successfully synthesized the single-phase polycrystalline samples of $Sn_{1-x}Ag_xTe$ with a wide range of $x$ = 0-0.5 using the high-pressure synthesis method, while the solubility limit of Ag for the Sn site is $x$ < 0.2 for the synthesis at ambient-pressure condition. The crystal structure of $Sn_{1-x}Ag_xTe$ at room temperature was confirmed as the cubic NaCl-type structure, which did not vary upon Ag substitution. The lattice constant monotonically decreased with increasing Ag content. For $x$ = 0.15-0.25, bulk superconductivity with a $T_c$ of 2.4 K was observed, and the optimal Ag content was $x$ = 0.2. The magnetization-field characteristics suggested that the HP-$Sn_{0.8}Ag_{0.2}Te$ superconductor is a type-II superconductor. The upper critical field estimated using the WHH model was 0.72 T. The superconductivity phase diagram of $Sn_{1-x}Ag_xTe$ seemed to be different from that of the In-doped $Sn_{1-x}In_xTe$ system. Therefore, the $Sn_{1-x}Ag_xTe$ superconducting phase will be useful for understanding the superconductivity nature and mechanisms of the carrier-doped SnTe system. In addition, this study proposes the advantage of the high-pressure synthesis technique for exploring new superconductors related to topological crystalline insulators with the NaCl-type structure.

**Acknowledgment**

The authors thank A. Iyo and T. D. Matsuda for fruitful discussion. This work was partly supported by Grant-in-Aid for Scientific Research (25707031, 26600077, 15H05886).

*E-mail: mizugu@tmu.ac.jp

1) X. L. Qi, and S. C. Zhang, Rev. Mod. Phys. **83**, 1057 (2011).
2) C. L. Kane, and E. J. Mele, Phys. Rev. Lett. **95**, 146802 (2005).
3) Y. L. Chen, J. G. Analytis, J. H. Chu, Z. K. Liu, S. K. Mo, X. L. Qi, H. J. Zhang, D. H. Lu, X. Dai, Z. Fang, S. C. Zhang, I. R. Fisher, Z. Hussain, and Z. X. She, Science **325**, 178 (2009).
4) L. Fu, and C. L. Kane, Phys. Rev. Lett. **100**, 096407 (2008)
5) Y. Ando, and L. Fu, Annu. Rev. Condens. Matter Phys. **6**, 361 (2015).
6) Y. S. Hor, A. J. Williams, J. G. Checkelsky, P. Roushan, J. Seo, Q. Xu, H. W. Zandbergen, A. Yazdani, N. P. Ong, and R. J. Cava, Phys. Rev. Lett. **104**, 057001 (2010).
7) S. Sasaki, M. Kriener, K. Segawa, K. Yada, Y. Tanaka, M. Sato, and Y. Ando, Phys. Rev. Lett. 107, 217001 (2011).
8) M. Kriener, K. Segawa, Z. Ren, S. Sasaki, and Y. Ando, Phys. Rev. Lett. **106**, 127004 (2011).
9) L. Fu, Phys. Rev. Lett 106, 106802 (2011).
10) T. H. Hsieh, H. Lin, J. Liu, W. Duan, A. Bansil, and L. Fu, Nat. Commun. **3**, 982(2012).

11) Y. Tanaka, Z. Ren, T. Sato, K. Nakayama, S. Souma, T. Takahashi, K. Segawa, and Y. Ando, Nat. Phys. 8, 800 (2012).
12) M. P. Mathur, D. W. Deis, C. K. Jones, and W. J. Carr, Jr., J. Phys. Chem. Solids **34**, 183 (1973).
13) A. S. Erickson, J. H. Chu, M. F. Toney, T. H. Geballe, and I. R. Fisher, Phys. Rev. B **79**, 024520 (2009).
14) G. Balakrishnan, L. Bawden, S. Cavendish, and M. R. Lees, Phys. Rev. B **87**, 140507 (2013).
15) M. Novak, S. Sasaki, M. Kriener, K. Segawa, and Y. Ando, Phys. Rev. B **88**, 140502 (2013).
16) R. D. Zhong, J. A. Schneeloch, X. Y. Shi, Z. J. Xu, C. Zhang, J. M. Tranquada, Q. Li, and G. D. Gu, Phys. Rev. B **88**, 020505 (2013).
17) N. Haldolaarachchige, Q. Gibson, W. Xie, M. B. Nielsen, S. Kushwaha, and R. J. Cava, Phys. Rev. B **93**, 024520 (2016).
18) F. Izumi, and K. Momma, Solid State Phenom. **130**, 15 (2007).
19) A. Iyo, K. Hira, K. Tokiwa, Y. Tanaka, I. Hase, T. Yanagisawa, N. Takeshita, K. Kihou, C. H. Lee, P. M. Shirage, P. Raychaudhuri, and H. Eisaki, Supercond. Sci. Technol. **27**, 025005 (2014).
20) K. Miyazawa, K. Kihou, P. M. Shirage, C. H. Lee, H. Kito, H. Eisaki, and A. Iyo, J. Phys. Soc. Jpn. **78**, 034712 (2009).
21) S. Iimura, S. Matsuishi, H. Sato, T. Hanna, Y. Muraba, S. W. Kim, J. E. Kim, M. Takata, and H. Hosono, Nat. Commun. **3**, 943 (2012).
22) Y. Mizuguchi, T. Hiroi, J. Kajitani, H. Takatsu, H. Kadowaki, and O. Miura, J. Phys. Soc. Jpn. **83**, 053704 (2014).
23) S. Kuroiwa, H. Kawashima, H. Kinoshita, H. Okabe, and J. Akimitsu, Physica C **466**, 11 (2007).
24) N. R. Werthamer, E. Helfand, and P. C. Hohemberg, Phys. Rev. **147**, 295 (1966).
25) L. Pan, D. Bérardan, and N. Dragoe, J. Am. Chem. Soc. **135**, 4914 (2013).

Table I. The synthesis conditions and lattice constant (*a*) for the major phase (phase1) and the minor phase (phase2).

| Nominal *x* | Synthesis condition | *a* (Å) Phase1 | *a* (Å) Phase2 |
|---|---|---|---|
| AP-0 (SnTe) | Vacuum (melted) | 6.31766(8) | - |
| AP-0.1 | Vacuum (600ºC) | 6.2548(1) | - |
| AP-0.2 | Vacuum (600ºC) | 6.2365(2) | - |
| AP-0.3 | Vacuum (600ºC) | 6.2381(2) | 6.0804(3) |
| AP-0.4 | Vacuum (600ºC) | 6.2222(4) | 6.0794(8) |
| AP-0.5 | Vacuum (600ºC) | 6.2130(7) | 6.0799(9) |
| HP-0.15 | 2 GPa (500ºC) | 6.2108(4) | 6.2920(1) (minor) |
| HP-0.2 | 2 GPa (500ºC) | 6.17670(8) | - |
| HP-0.25 | 2 GPa (500ºC) | 6.1403(3) | - |
| HP-0.33 | 2 GPa (500ºC) | 6.1076(5) | - |
| HP-0.5 | 2 GPa (500ºC) | 6.0828(2) | - |

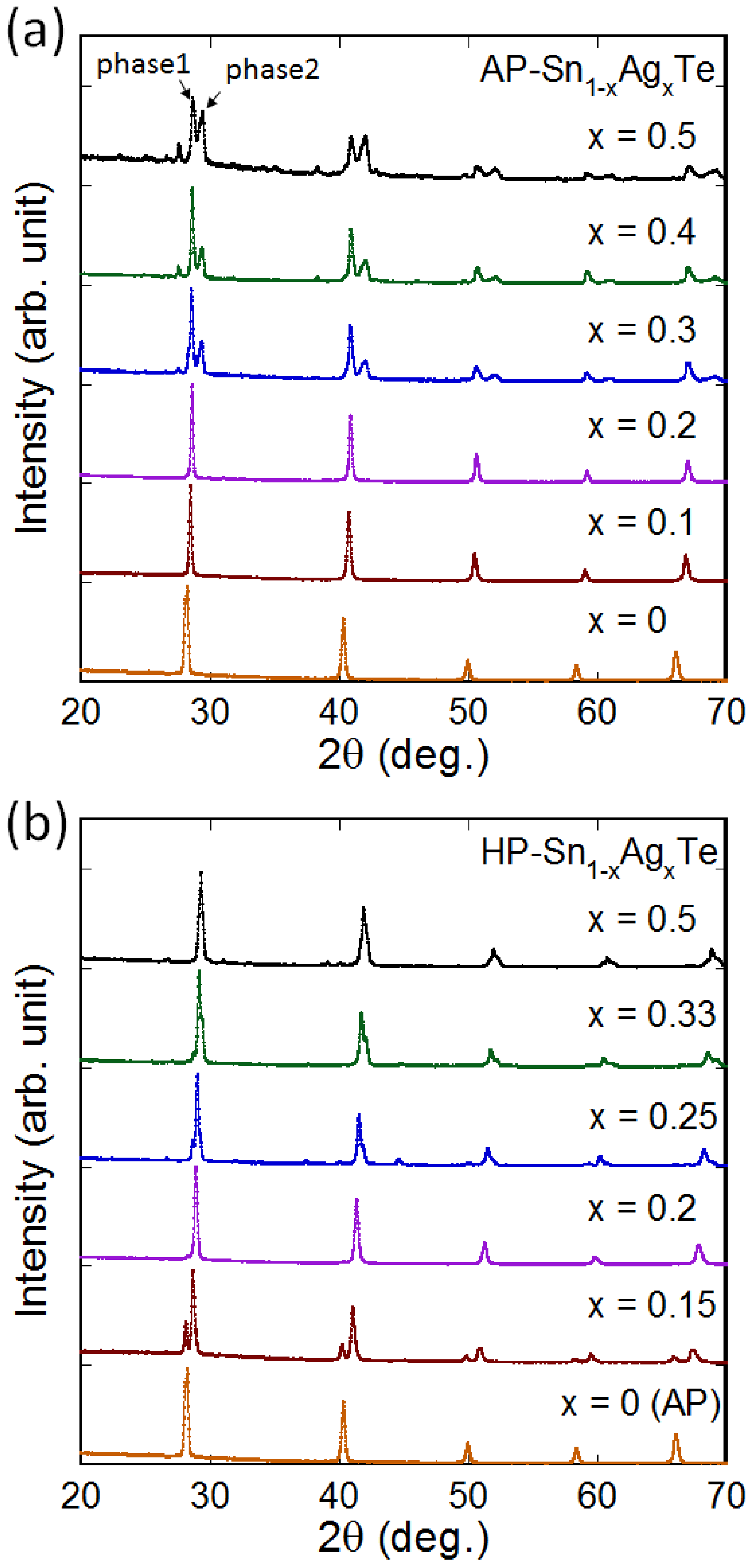

Figure 1. (a) XRD patterns for AP-$Sn_{1-x}Ag_xTe$. (b) XRD patterns for HP-$Sn_{1-x}Ag_xTe$.

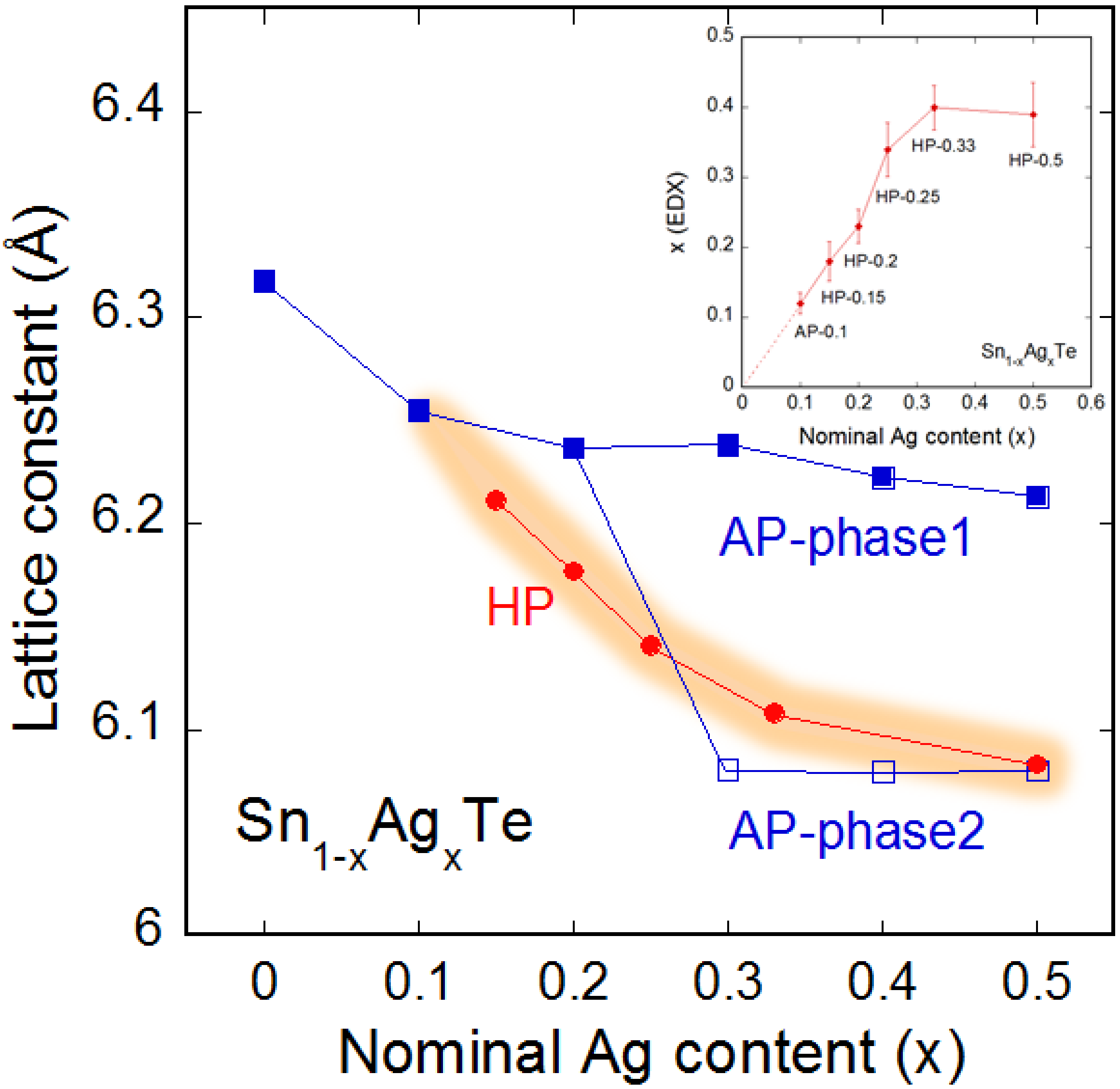

Figure 2. Nominal Ag content ($x$) dependences of lattice constant for AP- and HP- HP-$Sn_{1-x}Ag_xTe$. For AP-$Sn_{1-x}Ag_xTe$, phase separation into phase1 and phase2 is observed for $x$ = 0.3-0.5. The inset shows the Ag content analyzed using EDX as a function of nominal compositions for the single-phase samples.

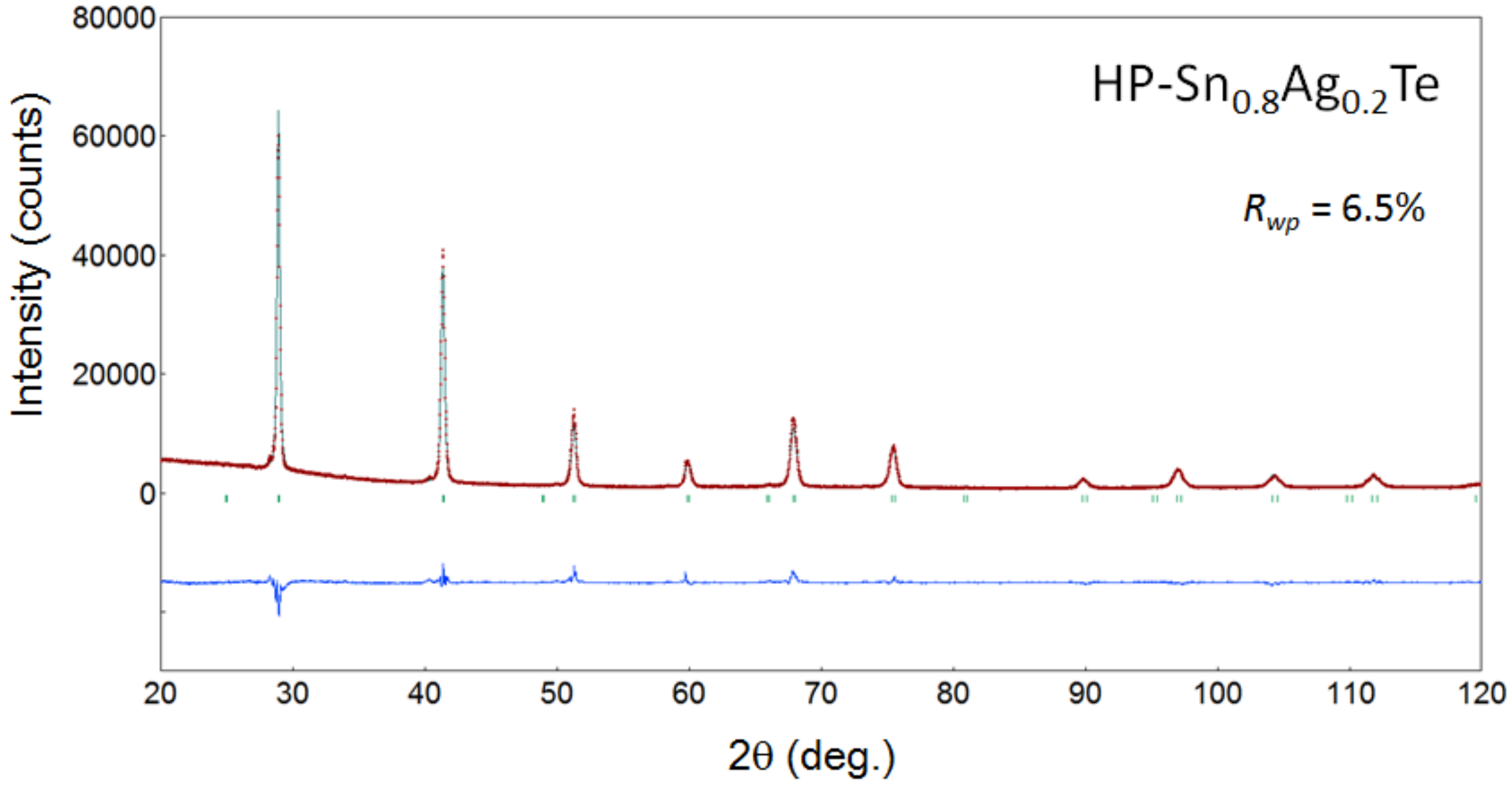

Figure 3. XRD pattern and fitting result of Rietveld refinement for HP-$Sn_{0.8}Ag_{0.2}Te$.

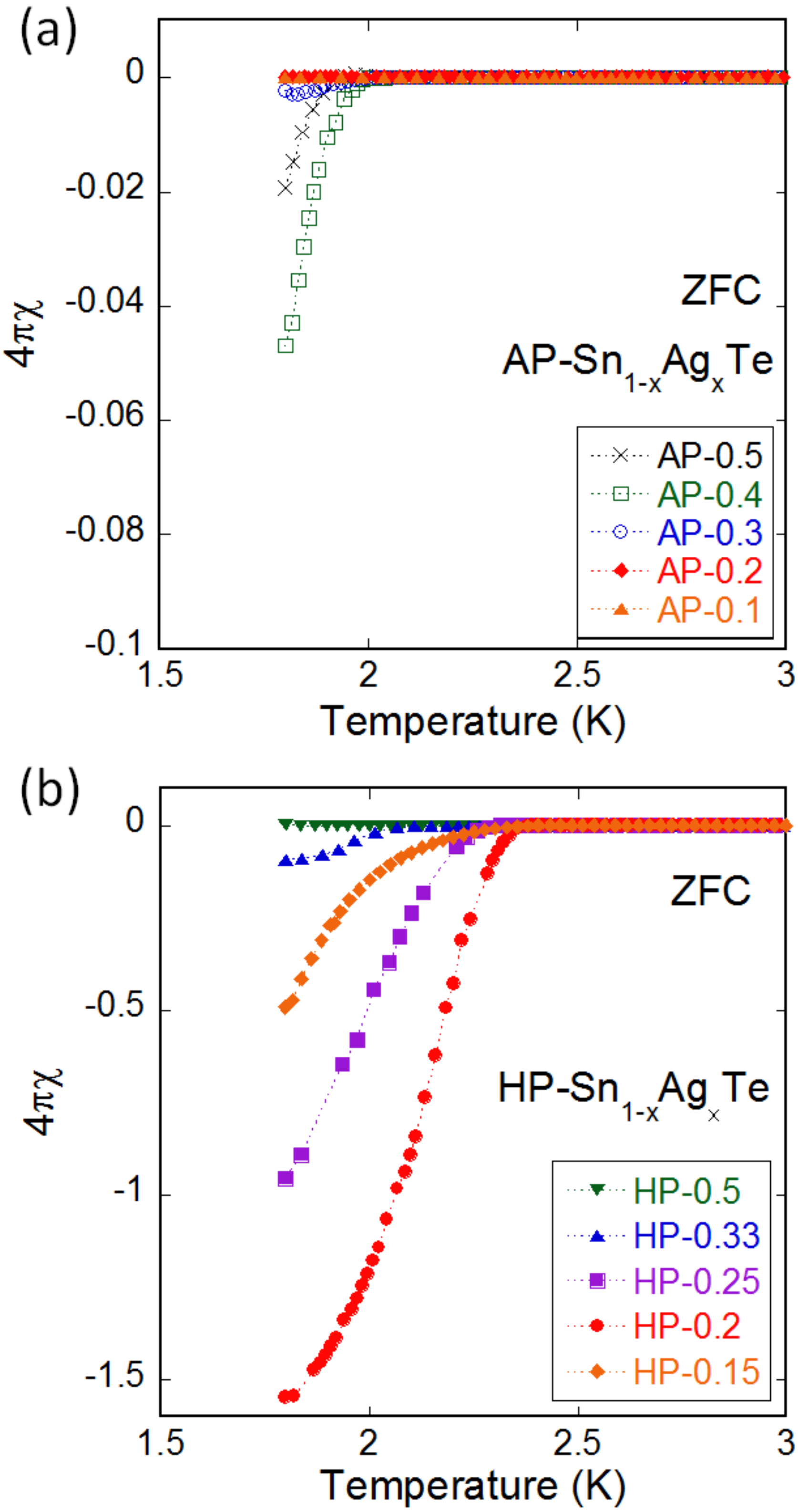

Figure 4. (a) Temperature dependences of $4\pi\chi$ (ZFC) for AP-$Sn_{1-x}Ag_xTe$. (b) Temperature dependences of $4\pi\chi$ (ZFC) for HP-$Sn_{1-x}Ag_xTe$.

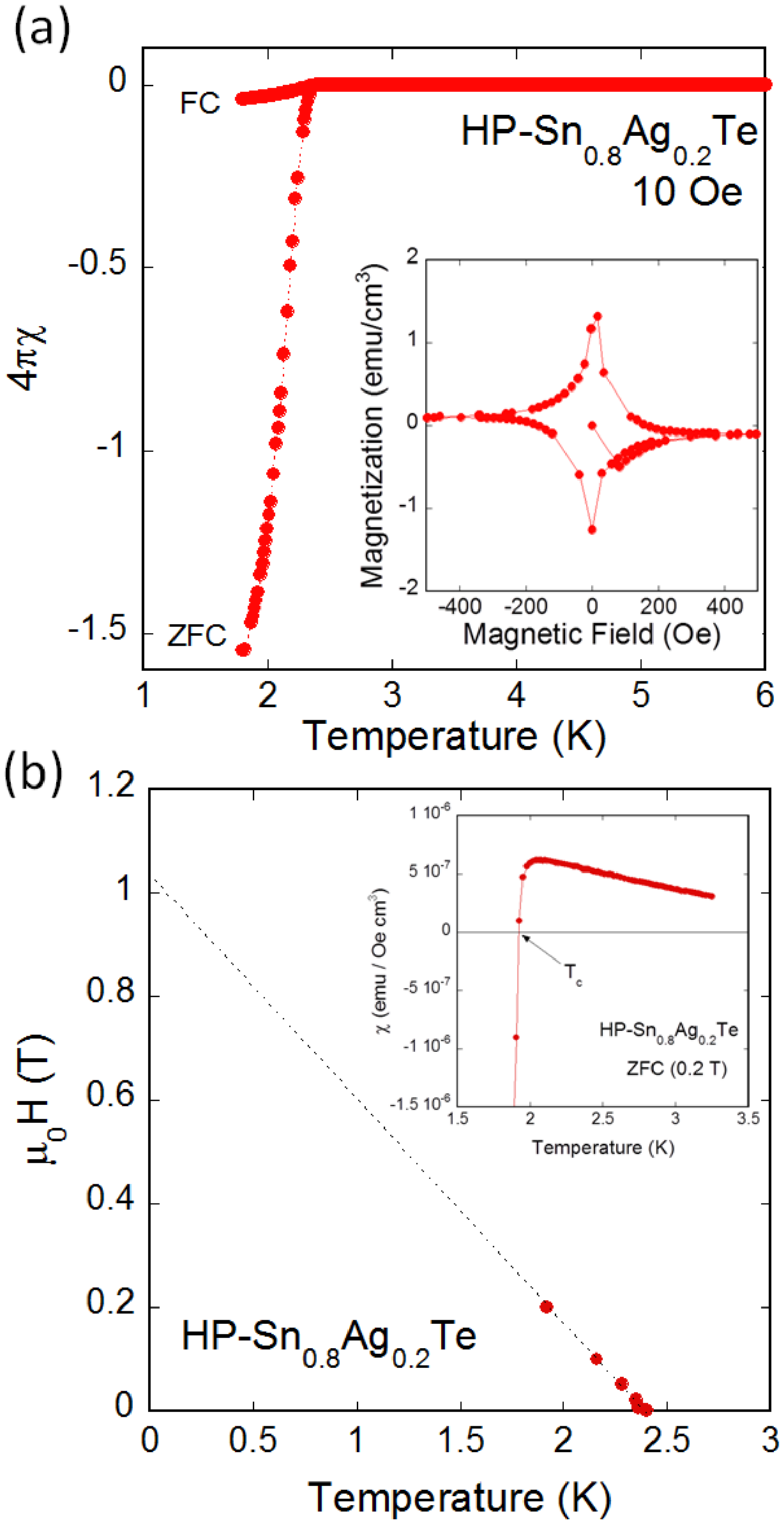

Figure 5. (a) Temperature dependences of $4\pi\chi$ (ZFC and FC) for HP-$Sn_{0.8}Ag_{0.2}Te$. The inset shows the magnetic field dependence of magnetization (*M*-*H* curve) for HP-$Sn_{0.8}Ag_{0.2}Te$. (b) Temperature dependences of $\mu_0 H_{c2}$ for HP-$Sn_{0.8}Ag_{0.2}Te$ estimated from the diamagnetic $T_c$. The diamagnetic $T_c$ under magnetic fields was defined to be the temperature at which $\chi$ became negative as shown in the inset (example at 0.2 T).

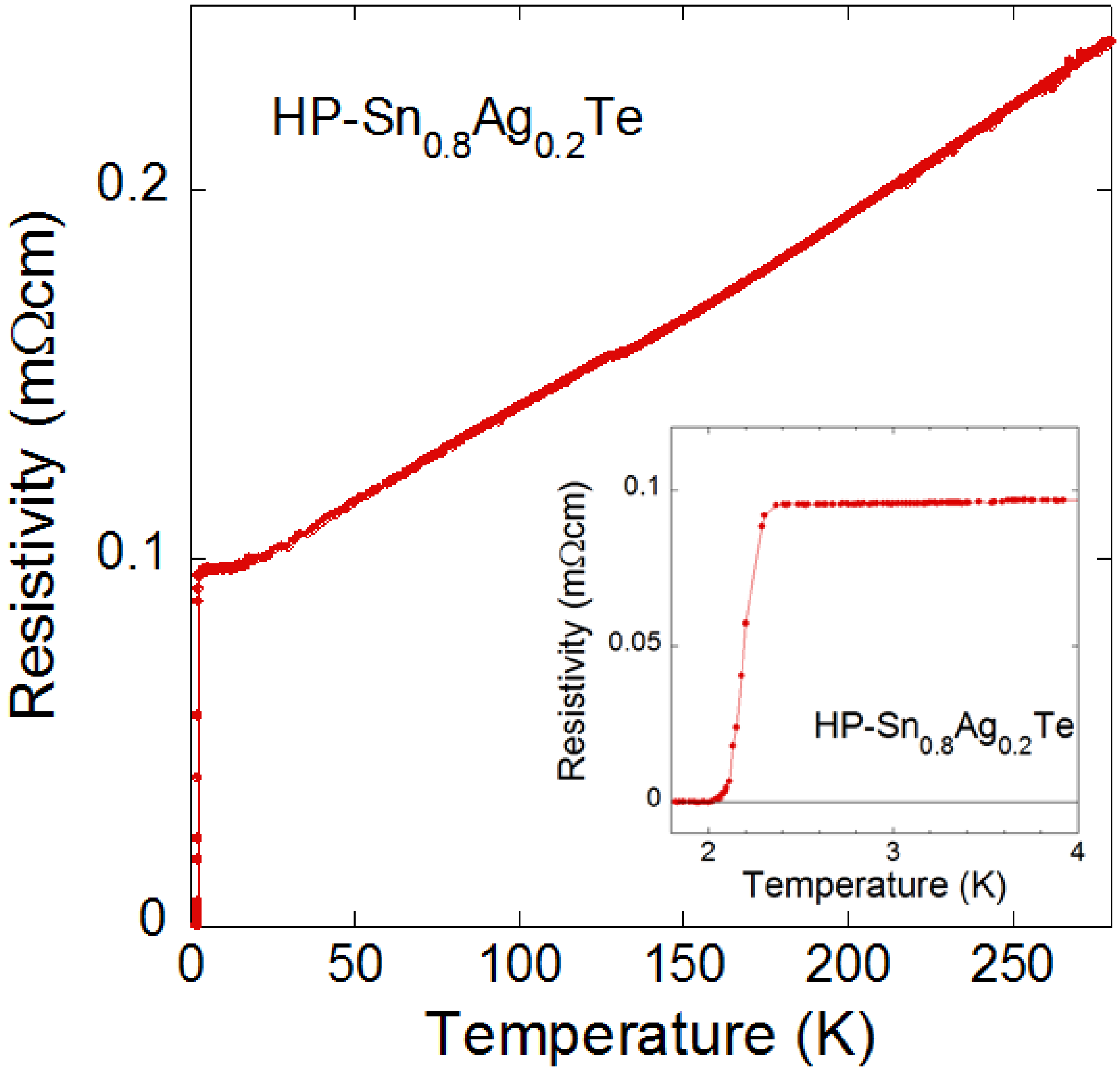

Figure 6. Temperature dependence of electrical resistivity for HP-$Sn_{0.8}Ag_{0.2}Te$. The inset shows the enlargement of the resistivity-temperature data at around the superconducting transition.

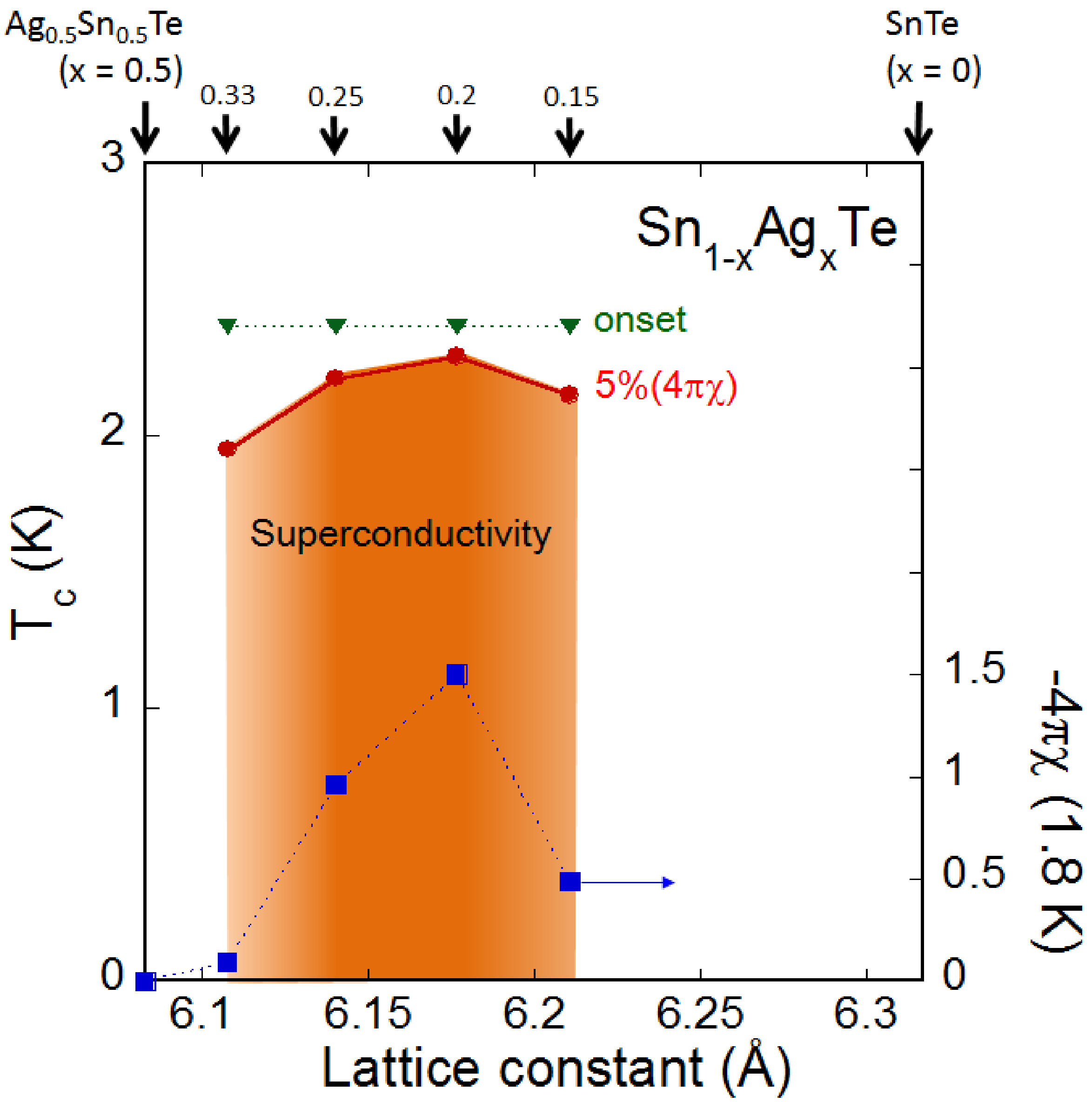

Figure 7. Transition temperatures ($T_c^{onset}$ and $T_c^{5\%(4\pi\chi)}$) for $Sn_{1-x}Ag_xTe$ as a function of lattice constant. To discuss the emergence of bulk superconductivity, $T_c^{5\%(4\pi\chi)}$ values were estimated as a temperature where shielding volume fraction exceeded 5%. The values of -4π$\chi$(ZFC) at 1.8 K are plotted as a function of lattice constant (left axis).